\documentclass[floatfix,aps,prd,twocolumn,notitlepage,shortbibliography,nofootinbib]{revtex4-1}
\pdfoutput=1
\usepackage[utf8]{inputenc}
\usepackage{amsmath}
\usepackage{amssymb}
\usepackage{braket}
\usepackage{graphicx}
\usepackage{bm}
\usepackage{hyperref}
\hypersetup{colorlinks=true,citecolor=Cyan,urlcolor=Red}
\usepackage[capitalise]{cleveref}
\crefname{equation}{}{}
\usepackage{microtype}

\newcommand{\zc}{\ensuremath{\overline{z}}}
\newcommand{\stirling}[2]{
    S_{#1}^{#2}
}
\newcommand{\dispSmatrix}{\ensuremath{%
        \mathcal{S}(ea + \xi)
    }
}
\DeclareMathOperator{\sinc}{sinc}
\DeclareMathOperator{\tr}{tr}

\usepackage[dvipsnames]{xcolor}
\newcommand{\be}{\begin{equation}}
\newcommand{\ee}{\end{equation}}
\newcommand{\bi}{\begin{itemize}}
\newcommand{\ei}{\end{itemize}}
\newcommand{\bea}{\begin{eqnarray}}
\newcommand{\eea}{\end{eqnarray}}
\newcommand{\ud}{\mathrm{d}}		
\usepackage{bbm}
\newcommand{\Eins}{\mathbbmss{1}}

\begin{document}

\title{Back-reaction in strong field QED: a toy model}
\author{Robin Ekman}
\email{robin.ekman@plymouth.ac.uk}
\author{Anton Ilderton}
\email{anton.ilderton@plymouth.ac.uk}
\affiliation{Centre for Mathematical Sciences, University of Plymouth, Plymouth, PL4 8AA, UK}

\begin{abstract}
    As a toy model for QED in strong background fields, we consider the impact of back-reaction and loop effects on scattering processes in quantum optics.
    We show that neglecting back-reaction misses qualitative and quantitative features of strong-field physics.
    We are able to study an analogue of the Narozhny-Ritus conjecture on the scaling of higher loop diagrams with intensity: we prove that there is no corresponding behaviour in our model.
    Implications for QED are identified and discussed.
\end{abstract}

\maketitle
\section{Introduction}
Dividing a system into a fixed background, and perturbations around it, is a standard and fruitful approach in many areas of physics.
The approach fails when the perturbation does not remain, in some sense, small compared to the background, and it then becomes necessary to account for ``back-reaction'' on the latter.

In the interaction of matter with intense lasers~\cite{ritus1985quantum,Dunne2009,RevModPhys.84.1177,king2016measuring}, one usually treats the laser fields as a fixed, plane wave background.
The assumption that back-reaction on this field can be neglected may be expected to break down when significant energy is depleted from the laser via, for example, radiation~\cite{PhysRevLett.118.154803,Blackburn:2019rfv} or pair production~\cite{PhysRevLett.101.200403,Fedotov:2010ja,PhysRevD.90.025016,Kasper:2015cca}.
New methods of calculation are then required.

\begin{figure}[t!]
    \centering
    \includegraphics[width=0.75\columnwidth]{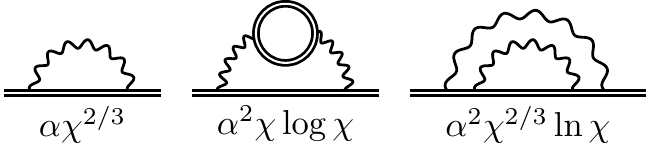}
    \caption{\label{fig:intro}
    Example QED loop diagrams in the Furry picture.
    Double lines indicate fermion propagators dressed to all orders by the background.
    The asymptotic scalings of the diagrams is also shown~[\citenum{Ritus:1970tf}, \citenum{Ritus:1972ky}, \citenum{Morozov:1975uah}] where the `quantum non\-linearity' parameter $\chi$ is the product of particle energy and external field intensity (see \cref{eq:chi-def}) and $\alpha$ is the fine-structure constant.
    }
\end{figure}

QED scattering in strong backgrounds is calculated in the `Furry picture' (background field perturbation theory)~\cite{Furry51,DeWitt:1967ub,tHooft:1975uxh,Boulware:1980av,Abbott:1981ke}.
Here the expansion parameter is the usual, small, coupling, but where the background is treated exactly at each order.
It has been conjectured, based on the scaling of certain higher loop diagrams in plane waves, see~\cref{fig:intro}, that the effective coupling parameter in the Furry picture is actually dependent on a (positive) power of the background field strength~\cite{PhysRevD.21.1176}.
If so, this would imply a breakdown of perturbative methods, or of the background field approximation, at sufficiently high intensity, and necessitate an all-orders resummation of Furry picture Feynman diagrams.
See~\cite{Fedotov_2017} for a review.

Investigating this conjectured behaviour is a target of future experiments~\cite{Blackburn:2018tsn,Baumann:2018ovl,Yakimenko:2018kih,DiPiazza:2019vwb}.
However, questions remain.
The conjecture is based on the special case of a constant `laser' field, and it is known that the associated scaling does not appear for general fields~\cite{Ilderton:2019vot}, nor does it hold at high energy~\cite{PhysRevD.99.076004,PhysRevD.99.085002}.

As these topics are demanding in full QED, we turn here to a toy model in an effort to shed some light on the problem.
This is the Jaynes-Cummings model~\cite{jaynes1963comparison} (JCM) of a single photon mode interacting with a single fermion spin.
The model has the advantages that it is exactly solvable, and that its three-point vertex mimics that of QED, allowing an analogy with Feynman diagrams.
While this is certainly a gross simplification of QED, single-mode models commonly reveal new physics and offer methods of including explicit quantum corrections which are otherwise hard to capture~\cite{berson1969electron,Bergou:1980cm,Bergou:1980cp,Heinzl:2018xnv}.
This approach will provide novel insights into both JCM and QED.

This paper is organised as follows.
In~\cref{SECT:JC} we review some relevant properties of JCM.
In \cref{sec:bgf-and-beyond} we show how the background field approximation, and corrections to it, arise in a systematic expansion.
We give an example of back-reaction at strong coupling in \cref{SECT:EG1}, showing that the well-known collapse/revival physics of JCM cannot be captured by \emph{any} ``higher loop" calculation unless back-reaction is also included.
In \cref{sec:furry}, we prove that the effective expansion parameter in the JCM Furry picture is just the coupling.
We show, however, that even for weak couplings it is necessary to include emissions as well as loop corrections in order to fully capture strong field physics.
We conclude in \cref{sec:conclusion}.

\section{The Jaynes-Cummings model}\label{SECT:JC}
The JCM Hamiltonian couples a single photon mode, frequency $\omega$, to a two-level system, energy gap $\omega_a$~\cite{jaynes1963comparison}
\begin{equation}
    H = \omega a^\dagger a + \omega_a \tau_3 + g a^\dagger \tau_- + g a \tau_+,
    \label{eq:JC-hamiltonian}
\end{equation}
where $a$ and $a^\dagger$ are the usual ladder operators for the photon mode, $[a, a^\dagger] = 1$, the operators $\tau$ obey the SU(2) algebra $\big[\tau_3, \tau_\pm \big] = \pm \tau_\pm$, $\big[\tau_+, \tau_- \big] = \tau_3$, and $g$ is the coupling.
For reviews see~\cite{Stenholm,knight1980rabi,GerryKnight}.
To aid the analogy with QED, we can take the two-level system to describe the spin states $\ket{\downarrow}$ and $\ket{\uparrow}$ of an electron, which requires $\omega_a=0$.
Calculations are however simpler, and our results equivalent, in the resonance limit $\omega_a=\omega$, which we adopt here.
The $\tau$ may be represented in terms of the spin states as
\begin{equation*}
    \tau_+ = \ket{\uparrow}\bra{\downarrow},
    \quad
    \tau_- = \ket{\downarrow}\bra{\uparrow},
    \quad
    \tau_3 = \frac{1}{2} \ket{\uparrow}\bra{\uparrow} - \frac{1}{2} \ket{\downarrow}\bra{\downarrow}
    .
\end{equation*}
We are interested in scattering.
The time evolution operator in the interaction picture $U(t)$ may be written down exactly (the Dyson series is convergent for all $g$); acting on an initial state which is, for simplicity, spin down but with arbitrary photon content `in'~\cite{Stenholm},
\begin{equation}
    \label{Udef}
    U(t) \ket{\textnormal{in}, \downarrow} =
        \cos gt \sqrt{\hat{n}} \ket{\text{in}, \downarrow}
        - i a \frac{\sin gt\sqrt{\hat{n}}}{\sqrt{\hat{n}}}
        \ket{\textnormal{in}, \uparrow}
        \;,
\end{equation}
where $\hat{n} = a^\dagger a$ is the photon number operator.
(Functions of operators are defined by their power series.)
As $U$ is a function of $gt$, JCM always runs to a strongly coupled regime as time evolves.
If $g$ is asymptotically switched, we obtain the ``$S$-matrix'' $\mathcal{S}$ by replacing $gt$ in \cref{Udef} with the integral of $g(t)$ over all time; this defines the dimensionless coupling $e$ which mimics the charge in QED.

Since the interaction $H_I$ is a three-point vertex which couples a single photon to the spin, we can draw Feynman diagrams which are analogous to those of QED for JCM processes.
For example, consider the analogue of Compton scattering;
\begin{multline}
    \bra{1,\downarrow}\mathcal{S}\ket{1,\downarrow} - 1
    = \cos(e) - 1
    = -\frac{e^2}{2!} + \frac{e^4}{4!} + \ldots
    \\
    =
    \raisebox{0pt}{\includegraphics{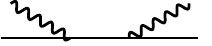}}
    +
    \raisebox{0pt}{\includegraphics{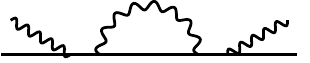}}
    +
    \ldots
\end{multline}
The `$-1$' above subtracts the disconnected contribution.
The $\mathcal{O}(e^2)$ term corresponds to tree level Compton scattering; this is easily verified by expanding \cref{Udef} in powers of $e$, and checking that this term comes from contractions only between $a$-operators in $\mathcal{S}$ and $a$-operators in the asymptotic states.
The $\mathcal{O}(e^4)$ term, on the other hand, also contains contractions of $a$-operators within the $S$-matrix, which correspond to one-loop corrections (of which we show only one of the possible diagrams).

\begin{figure}[t!]
    \includegraphics[width=0.45\textwidth]{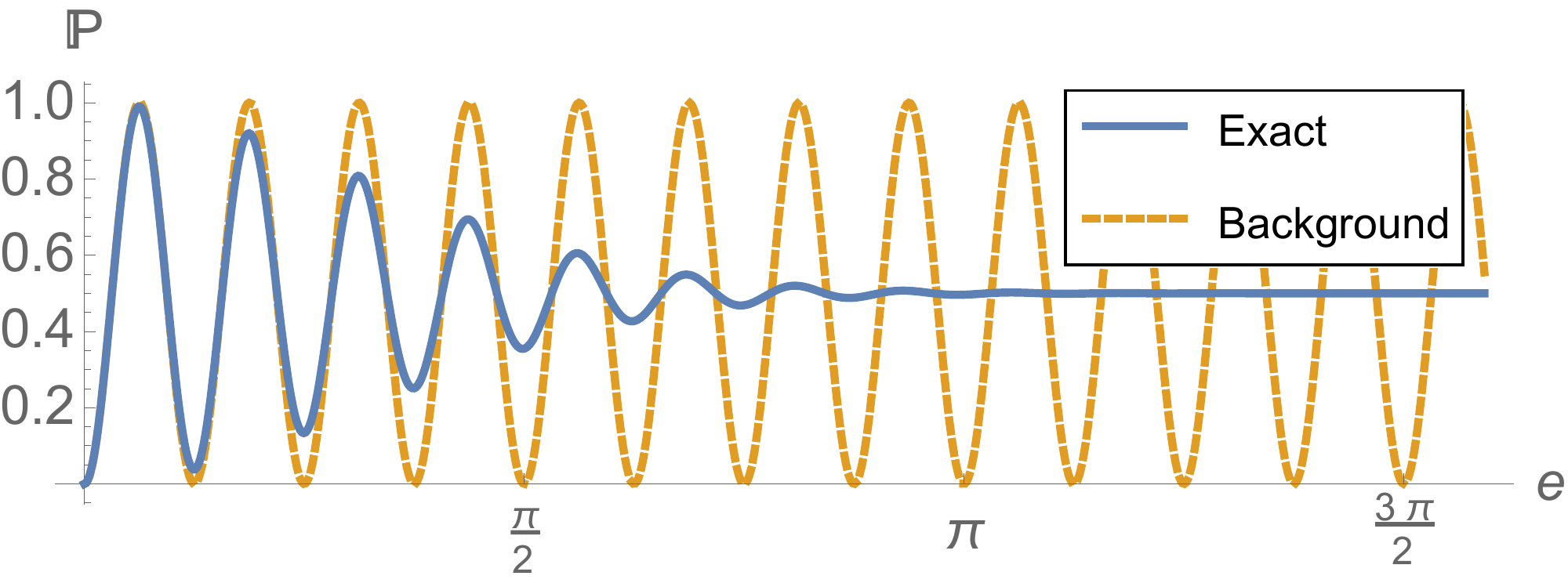}
    \qquad\qquad
    \includegraphics[width=0.45\textwidth]{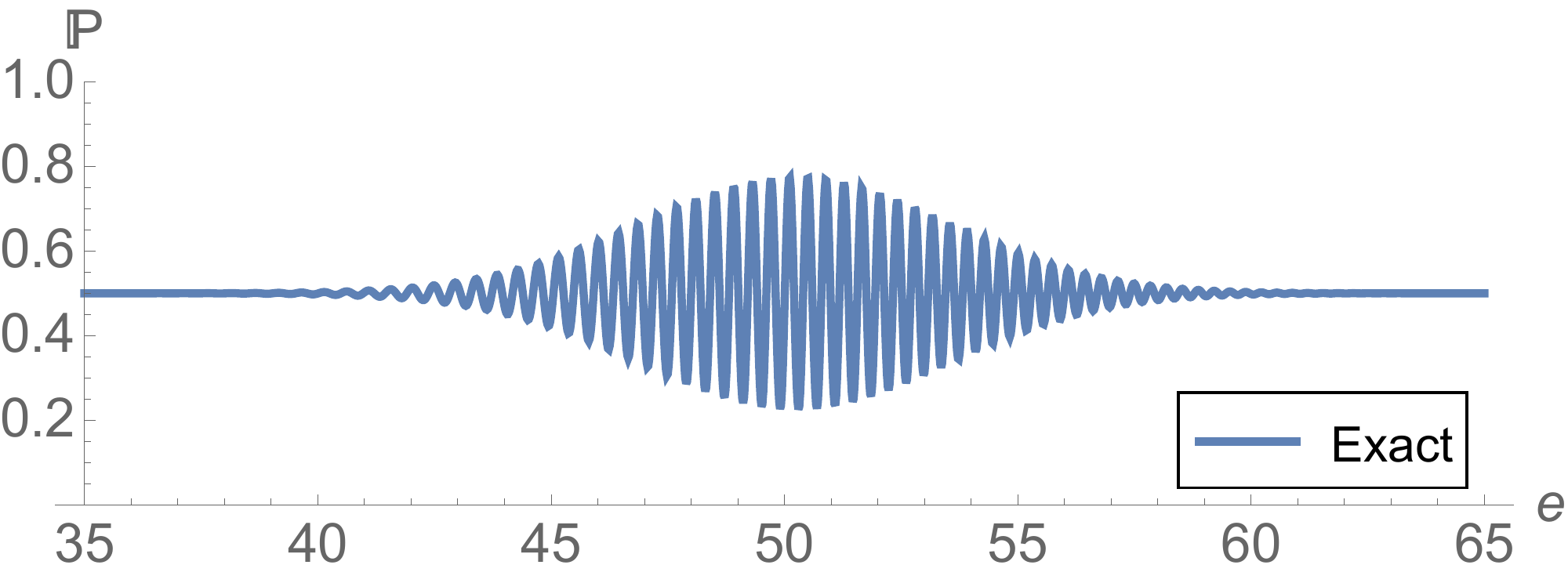}
    \caption{%
        \label{TNG}
        Comparison of the spin flip probability (blue/solid) with that given by the background field approximation (yellow/dashed) for $z = 8$.
        The latter (shown only in the upper panel for clarity) captures only the initial Rabi oscillations, which it exhibits for all $e$, whereas the exact results show collapse (upper panel) and revivals (lower panel).
    }
\end{figure}

\subsection{Motivation: spin flip and collapse/revivals}
\label{sec:motivation}

For comparison with strong field QED~\cite{ritus1985quantum,Dunne2009,RevModPhys.84.1177,king2016measuring}, we need a (strong) background modelling the laser.
We therefore review here relevant results of JCM with coherent states, which are closely related to background fields~\cite{KibbleFreq,Frantz,Gavrilov:1990qa}.

Consider placing a spin down particle in a coherent state $\ket{z}$ of photons, and asking for the probability that the spin flips as time evolves, which, from above, \cref{Udef}, is the same as increasing the coupling $e$.
In terms of the displacement operator $D(z) = \exp (z a^\dagger - {\bar z} a)$ the initial state is
\begin{equation}
\ket{z,\downarrow} \equiv     D(z) \ket{0,\downarrow}
    .
\end{equation}
Acting with the $S$-matrix, projecting onto the spin-up state, and summing over all possible final photon states, we obtain the total, inclusive, probability of spin flip,
\begin{equation}
    \label{def:flip}
    \begin{split}
        \mathbb{P} &:=
        \sum\limits_{n=0}^\infty |\bra{n,\uparrow}\mathcal{S}\ket{z,\downarrow}|^2 \\
                   &=
                   \frac{1}{2} - \frac{1}{2} \mathrm{e}^{-|z|^2} \sum\limits_{r=0}^\infty \frac{|z|^{2r}}{r!} \cos 2e\sqrt{r}.
    \end{split}
\end{equation}
The properties of the spin flip probability are well known.
As time evolves, $\mathbb{P}$ initially exhibits regular Rabi oscillations with frequency $e|z|$, before `collapsing' to $\mathbb{P} = 1/2$ \cite{PhysRev.140.A1051}, as illustrated in~\cref{TNG}.
The collapse is complete before $e\simeq \pi$, and the probability remains at $1/2$ until later times when it `revives' at $e = 2 \pi |z|$, collapses, and revives again at $e = 2\pi n |z| $ for $n=1, 2, 3 \ldots$~\cite{PhysRevLett.44.1323}.

Let us compare this behaviour with that predicted by the background field approximation, which is given replacing $\{a,a^\dagger\} \to \{z,z^\dagger\}$, in the Hamiltonian, so that the electromagnetic field is fixed and classical, and the state space is spanned by $\ket{\uparrow}, \ket{\downarrow}$.
One finds
\begin{equation}
    \label{dave}
    \mathbb{P} \simeq  \sin^2(e |z|)
    .
\end{equation}
Now, this approximation would correspond, in QED, to performing a tree level calculation of the exclusive (i.e.~no emission) spin flip probability in a background, which neglects both quantum corrections (loops) and back-reaction on the background through emission.
Hence, if the approximation \cref{dave} differs from the exact result \cref{def:flip}, then either quantum corrections, back-reaction, or both, are important.
The two results \cref{def:flip} and \cref{dave} are compared in \cref{TNG}.
For small coupling $e$, the tree level background field approximation works well, reproducing the Rabi oscillations, which are essentially classical.
However, the oscillations are of fixed amplitude for all $e$: the approximation does not reproduce the collapse or revival of the probability, for which it follows that loop corrections and/or emission are relevant.
We will identify which in \cref{SECT:EG1}.

\section{The background field approximation and beyond}
\label{sec:bgf-and-beyond}

We consider again the initial state $\ket{z,\downarrow}$.
Acting with the $S$-matrix, our aim is to write the evolved state $\mathcal{S}\ket{z,\downarrow}$ in such a way that the background field approximation and corrections to it become explicit.

We begin with the well-known property of displacement operators that, for any function $f$ of $a$ and $a^\dagger$,
\begin{equation}
    \label{disp-vanligt}
	D^\dagger(z) f(a , a^\dagger) D(z) = f(a + z, a^\dagger + \bar{z}).
\end{equation}
If we choose $f$ to be the $S$-matrix, it follows from \cref{disp-vanligt} that any amplitude between coherent states is equivalent to an amplitude in the presence of a background $z$ of the $a$-modes~\cite{KibbleFreq,Frantz,Gavrilov:1990qa} (where the matter content of both amplitudes is the same).
We observe that the $S$-matrix \cref{Udef} is a function of $e a$;
it is useful to make this dependence explicit, writing $\mathcal{S}\equiv \mathcal{S}(ea)$.
We also write unity in the space of $a$-modes, $\Eins_a$, as
\begin{equation}
    \label{EINS-DEF}
	\Eins_a = \sum\limits_n  D(z) \ket{n}\bra{n} D^\dagger(z) =: \sum\limits_n  \ket{z_n}\bra{z_n}
.
\end{equation}
The $\ket{z_n}$ are ``displaced number states'' (and $\ket{z_0} \equiv \ket{z}$  above).
(See Refs.~\cite{PhysRevA.41.2645,Nieto:1996eh} for reviews, \cite{Heinzl:2018xnv} and references therein for their use in other single-mode approximations to QFT problems.)
Working with these states instead of the number states $\ket{n}$ simply corresponds to using a basis which makes \emph{explicit} reference to the initial coherent state, $z$, in the system.

Combining \cref{EINS-DEF} and \cref{disp-vanligt} and defining $\xi := e z$ we have
\begin{equation}
    \label{precis-allt}
    \begin{split}
        \mathcal{S}(ea) \ket{z, \downarrow} &\equiv  \Eins_a \mathcal{S}(ea) D(z) \ket{0, \downarrow} \\
        &= \sum\limits_n \ket{z_n}\bra{n} \mathcal{S}(ea+\xi) \ket{0, \downarrow}
        \;.
    \end{split}
\end{equation}
This expression implicitly identifies the two truncations which lie behind the background field approximation and corrections to it.

\begin{itemize}
    \item[i)] Truncating the sum over $n$ in \cref{precis-allt} turns $\Eins_a$ into a projection operator which limits the possible \emph{final} state space of the photonic modes.
        This does not, though, prevent photons being created and destroyed during scattering (loop effects) which brings us to the second approximation.
    \item[ii)] The evolution operator in \cref{precis-allt} has become a function of $ea+\xi$.
        The dimensionless coupling to the background, and the analogue of the dimensionless intensity parameter in QED~\cite{ritus1985quantum,RevModPhys.84.1177} is $a_0:=|\xi|$.
        Expanding $\mathcal{S}$ in powers of $e$ at fixed~$\xi$ gives, at leading order, the semiclassical (tree level) approximation to each of the amplitudes $\bra{n}\mathcal{S}(ea+\xi)\ket{0,\downarrow}$ plus an infinite series of quantum corrections; this expansion in powers of $e$ corresponds to the Furry picture loop expansion in QED, in which the coupling to the background, $a_0$, is treated exactly~\cite{Furry51}.
\end{itemize}

To illustrate, consider the most severe truncation, which limits the sum over final states to $n \in \{0\}$ and also retains only the zeroth order terms in $e$.
Then we have the approximation
\begin{equation}
    \label{amp-0-0}
	\mathcal{S}(ea) \ket{z,\downarrow} \simeq \ket{z}  \bra{0} \mathcal{S}(\xi) \ket{0,\downarrow}
\end{equation}
in which $\mathcal{S}(\xi)$ acts only on the spin degrees of freedom, and where the final photon state is forced to be exactly equal to the initial state; the photons are spectator modes, and the spin degrees of freedom are affected only by the external field $z$.
The amplitude in \cref{amp-0-0} then corresponds to the tree level two-point function in a background;
the spin flips in an external field, or it does not.
The corresponding approximation to the total flip probability is given by \cref{dave} with $e|z| = a_0$, reproducing the Rabi oscillations.
(In this approximation the inclusive flip probability is exactly equal to the exclusive probability.)

We can add quantum corrections, corresponding to loops in QED, by retaining higher orders in $e$.
Retaining all orders in $e$ yields an all-loops result.
In the following two Sections we consider two examples in which loop effects and back-reaction impact the physics of JCM.

\section{back-reaction at strong coupling}
\label{SECT:EG1}
Investigations of the high-intensity behaviour of scattering processes in QED have focussed on the addition of higher loop corrections, and how they scale.
For our first example we therefore consider adding loop corrections to the tree-level background field calculation of spin flip (Rabi oscillations) in JCM, to see if the collapse and revivals are recovered.
It is convenient here to work with the variables $e$ and $z$ of JCM; this will allow us to study the impact of back-reaction at strong coupling $e\gg1$.
(We consider the Furry expansion proper in the next section.)

\begin{figure}[t!]
    \includegraphics[width=0.45\textwidth]{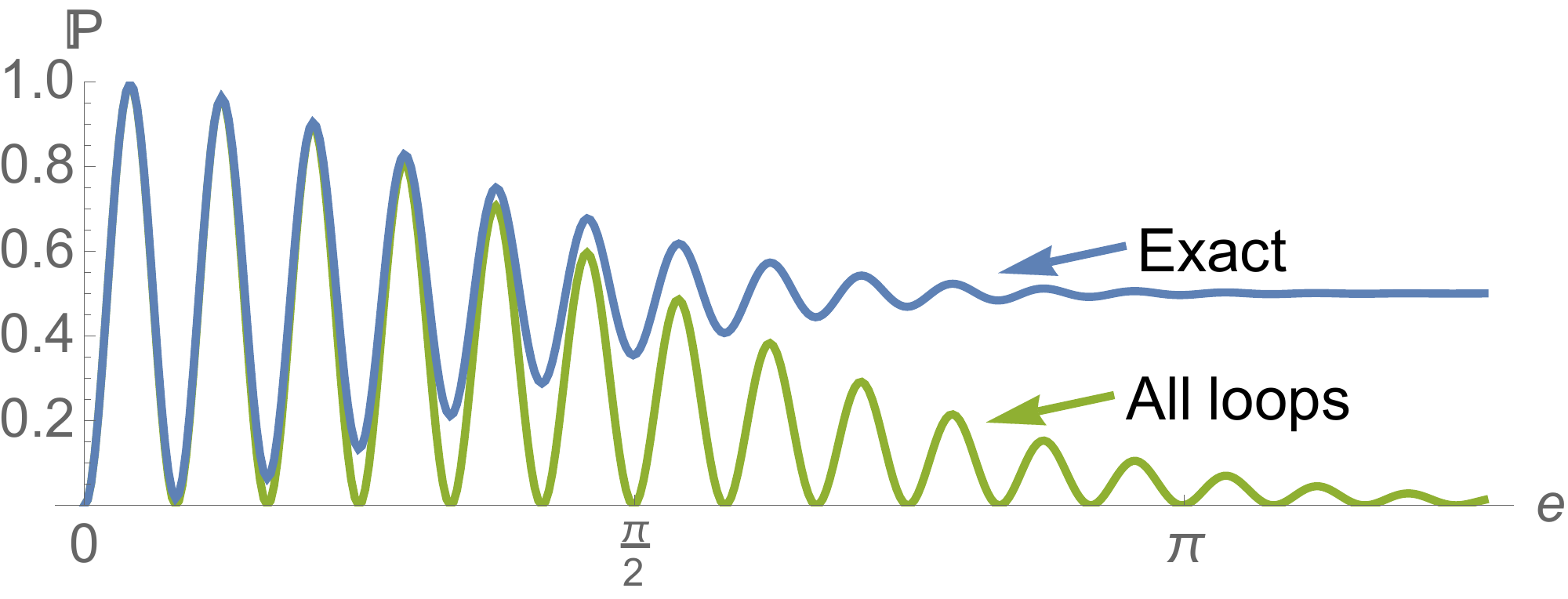}
    \includegraphics[width=0.45\textwidth]{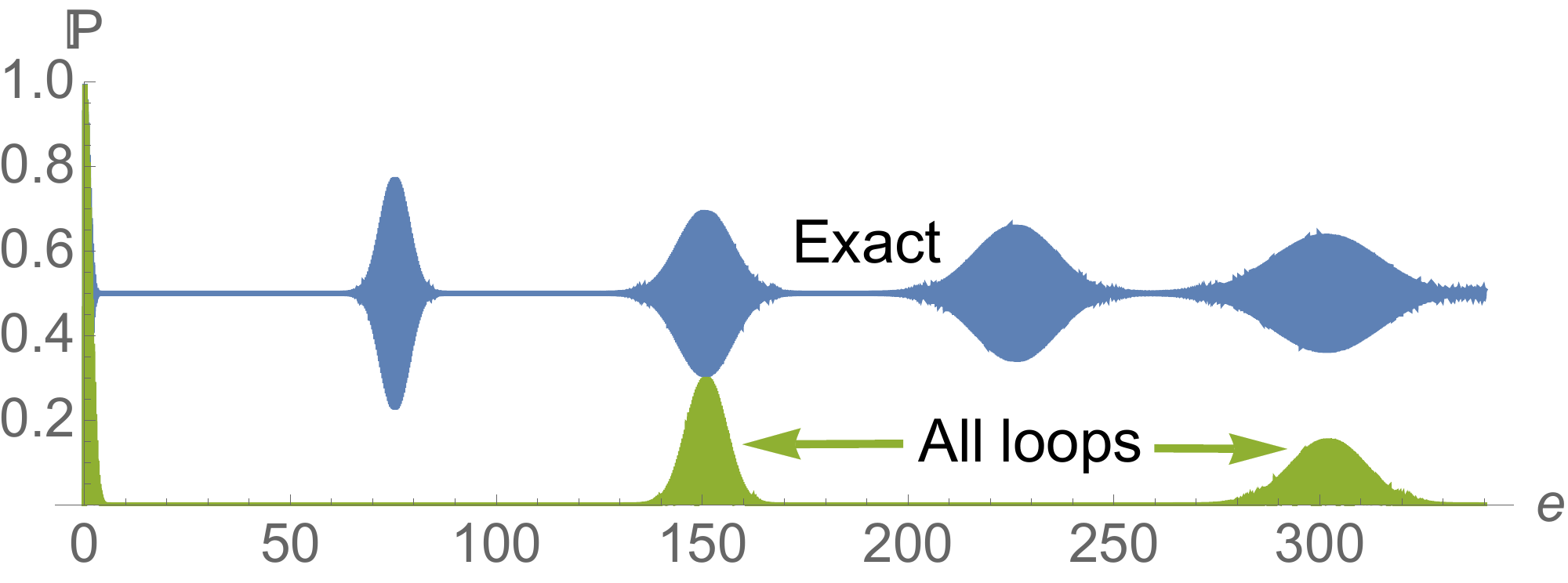}
    \caption{%
        \label{fig:no-revival}
        Collapse and revivals, with $z=12$; adding all loop corrections to the tree-level result kills the Rabi oscillations, but only partially recovers collapse and revivals -- in particular the collapse is to 0 rather than $1/2$, and all odd-numbered revivals are missed.
    }
\end{figure}

We truncate to $n = 0$ in \cref{precis-allt}, which neglects all emissions, but make no other approximation, hence all loops are retained.
The evolved state is then
\begin{equation}
        \mathcal{S}(ea) \ket{z, \downarrow} \simeq \ket{z}\bra{0} \mathcal{S}(ea+\xi) \ket{0, \downarrow}
        \;,
\end{equation}
and the \emph{total} spin-flip probability is
\begin{equation}
    \begin{split}
        \mathbb{P}
        &\simeq |\bra{0,\uparrow} \mathcal{S}(ea+\xi) \ket{0, \downarrow}|^2
               \\
               & =   |\bra{z,\uparrow} \mathcal{S}(ea) \ket{z, \downarrow}|^2
               \\
               & = |z^2|e^{-2|z|^2} \bigg|\sum\limits_{n=0}^\infty
               \frac{|z|^{2n}}{n!}\frac{\sin e\sqrt{n+1}}{\sqrt{n+1}}\bigg|^2 \;.
    \end{split}
    \label{eq:exact-prob}
\end{equation}
The second line of the above, written in terms of the original coherent state, emphasises that this is still a background field approximation in the sense that the final and initial photon states are the same.
The result is plotted in \cref{fig:no-revival} along with the exact result \cref{def:flip}.
The all-loops result shows a collapse and revivals: however, the collapse is to $0$ rather than to $1/2$, and the first revival in the exact result is missed entirely.
In fact, all odd-numbered revivals are missed.

\begin{figure}[]
    \centering
    \includegraphics[width=\linewidth]{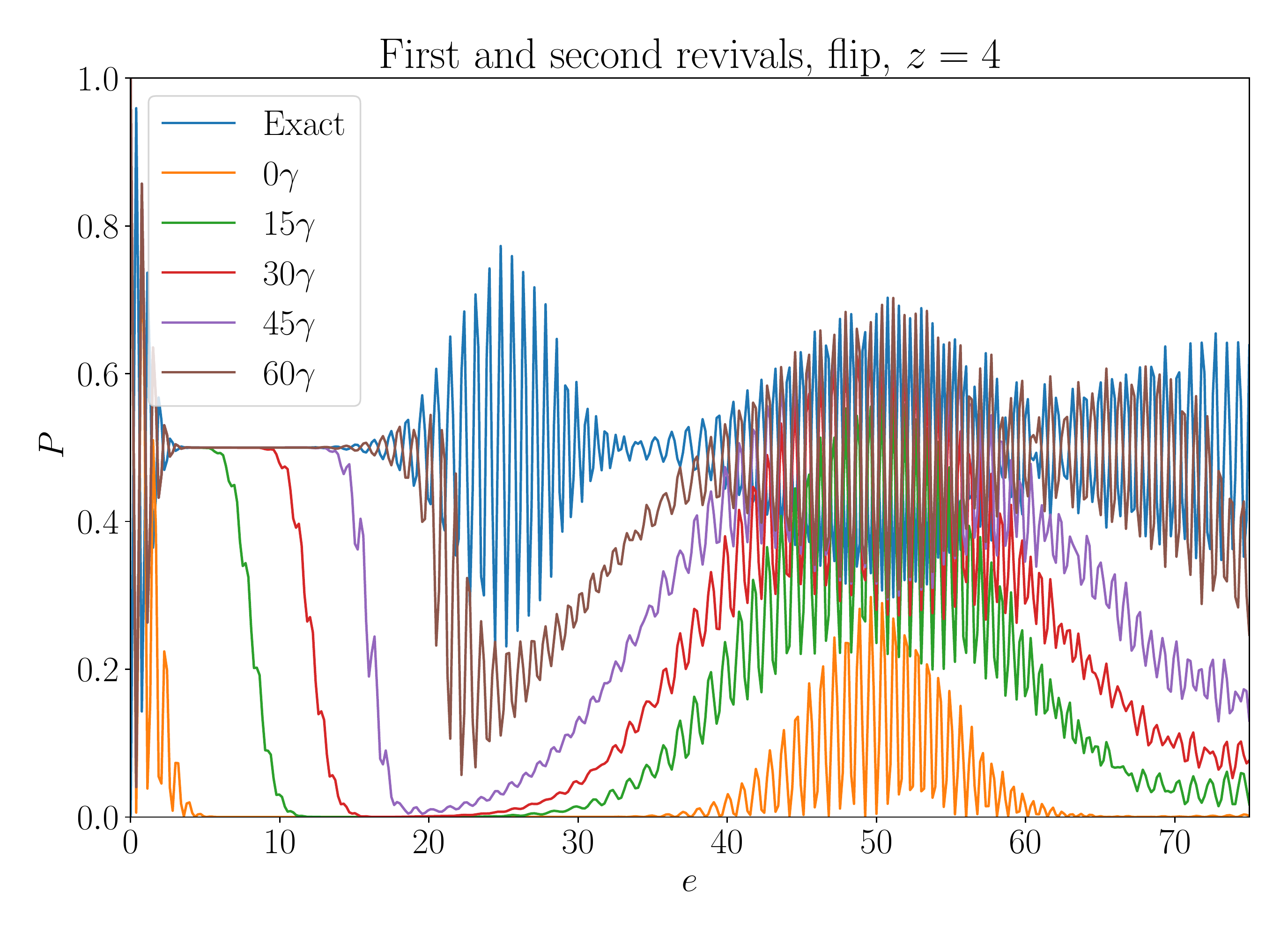}
    \caption{
        \label{fig:second-revival}
        Spin flip probability at fixed $|z|$ and varying coupling.
        ``Exact'' refers to the probability given by \cref{eq:exact-prob}, and $N\gamma$ means up to $N$ emissions included.
        Clearly, the first revival is missed unless $N$ is quite large, while the second revival is seen even with $N = 0$, although quantitatively incorrect.
        }
\end{figure}

Thus we learn that even adding \emph{all} loop corrections in the background field approximation is not enough to capture the physics of JCM.
While loops contribute to the even-numbered revivals, the odd-numbered revivals must be driven by emission (back-reaction) on the initially coherent photon state.
Including these emissions, we can consider the partially inclusive probability $\mathbb{P}_N$ of spin flip with \emph{up to} $N$ photon emissions, which means summing $n$ in \cref{precis-allt} from 1 to $N$, with the result
\begin{equation}
    \mathbb{P}_N
    :=
    \sum_{n = 0}^N \big|\bra{n, \uparrow} \mathcal{S}(ea+\xi) \ket{0, \downarrow}\big|^2
    \;.
\end{equation}
The larger $N$ must be in order to give a good approximation of the inclusive flip probability, the more significant is back-reaction.
In \cref{fig:second-revival}, we plot $\mathbb{P}_N$ for $z=4$ and various $N$ up to $N=60$, along with the exact result.
The figure shows that a significant number of emissions are required in order to properly capture the collapse and revivals.

To understand these results we examine the time-evolved photon state.
The fact that the sums in the exact results are strongly peaked around $n = |z|^2$ allows us to use the well-known approximation~\cite{GerryKnight,PhysRevLett.44.1323}
\begin{equation}
    \sqrt{n}
    = \sqrt{|z|^2 +(n-|z|^2)}
      \simeq \frac{|z|}{2} + \frac{n}{2|z|}
      \;.
    \label{eq:quad}
    \end{equation}
    Using this, and assuming $|z|^2\gg1$, we find that the exact state may be approximately written as a `cat' superposition of coherent states~\cite{GerryKnight,PhysRevLett.44.1323},
    \begin{equation}
        \bra{\uparrow} S \ket{z, \downarrow}
        \simeq
        \frac{1}{2} \mathrm{e}^{-i |z|^2 \theta} \Ket{z\mathrm{e}^{-i\theta}}
        -\frac{1}{2} \mathrm{e}^{i |z|^2 \theta} \Ket{z\mathrm{e}^{i\theta}}
        \;,
        \label{eq:cat}
    \end{equation}
    with $\theta := e/(2|z|)$.
    It is helpful, in parallel with this analytic approximation,
to visualise the state via the Wigner quasiprobability distribution~\cite{PhysRev.40.749,PhysRev.177.1882,ferrynedjalkov2018}, widely used in quantum optics, and also in studies of pair production in QED~\cite{BialynickiBirula:1991tx,Hebenstreit:2010vz,PhysRevLett.109.100402,PhysRevD.83.065007}.
The Wigner function is defined as
\begin{equation}
    W(\gamma) = \tr \int \hat{\rho} \exp\left[
        \left(
            \zc (\gamma- a)
            - z (\overline{\gamma} - a^\dagger)
        \right)
    \right] \ud^2 z
    \;,
\end{equation}
where $\hat{\rho}$ is the reduced (i.e., traced over the spin) photon density matrix and $\gamma$ is a complex phase-space coordinate.
The Wigner function highlights (quantum) deviations from (classical) coherent states: the Wigner function of a coherent state $\ket{z}$ is a Gaussian centered at $\gamma = z$, and only coherent states and squeezed vacua have an everywhere non-negative Wigner function~\cite{hudson1974wigner}.
Interference in quantum superpositions of states appears in the Wigner function as regions of negative values~\cite{Kenfack2004negativity}.

\begin{figure}[t!]
    \centering
    \includegraphics[width=\columnwidth]{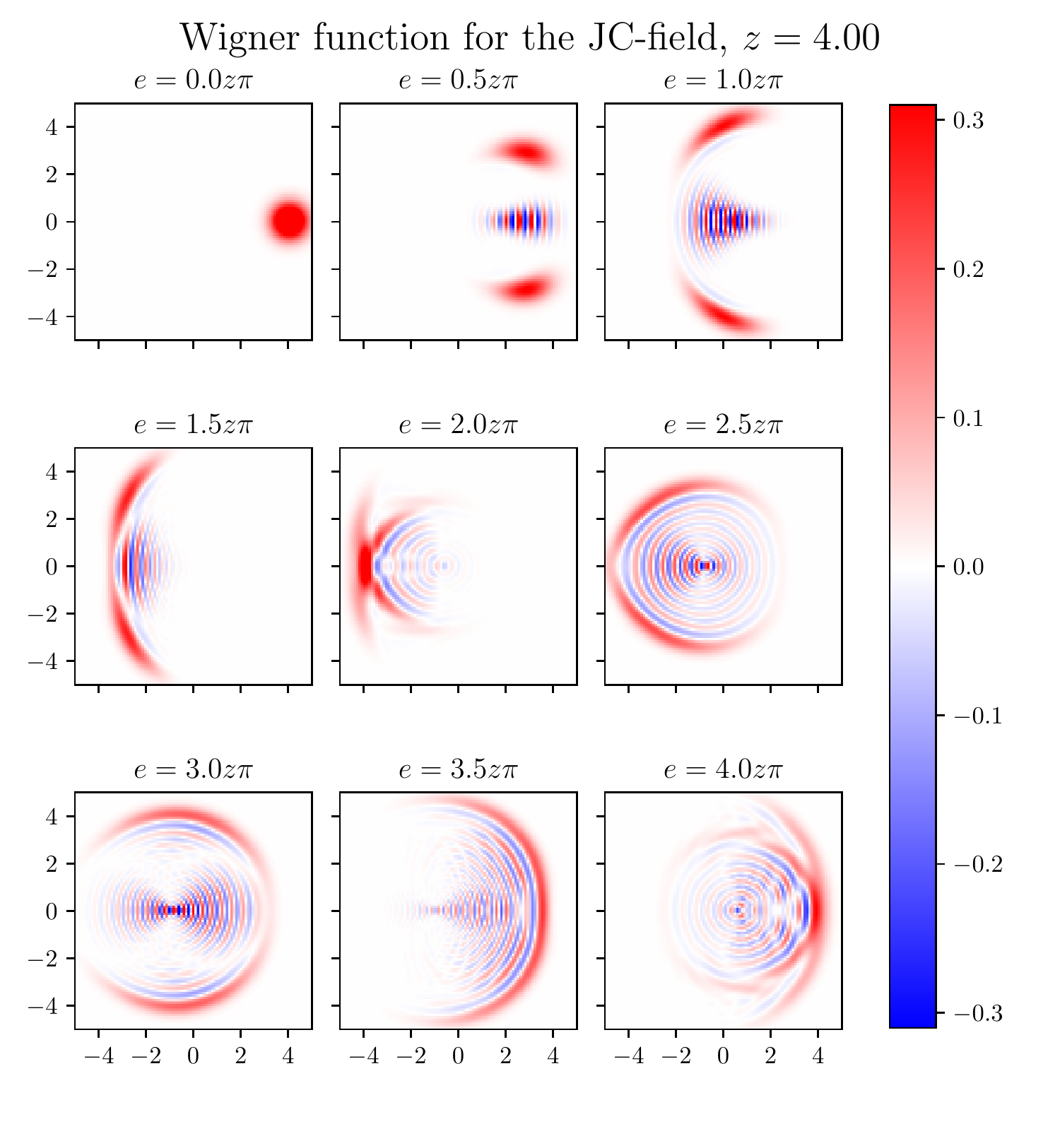}
    \caption{%
        Wigner function $W(\gamma)$ for an initially coherent state with $z = 4$, as a function of coupling/time.
        The horizontal/vertical axes are the real/imaginary parts of $\gamma$.
        The cat state, evident in the second and third panels, has been observed experimentally, and measurements of its Wigner function made~\cite{PhysRevLett.123.143605}.
        The quadratic terms neglected in (\ref{eq:quad}) lead to squeezing, and eventually the state completely loses coherence.
        Without back-reaction, $W(\gamma)$ would not change at all.
    }
    \label{fig:wigner}
\end{figure}

\cref{fig:wigner} shows the Wigner function for the evolved state in JCM at different times (couplings).
From this and \cref{eq:cat} we see that the initially coherent state splits into a superposition of two coherent states which counter-rotate in phase ($\gamma$) space as time evolves.
As they do, interference effects appear in the Wigner function.
The collapse is due to these non-classical effects in the `spinning' cat state~\cite{PhysRevA.43.346,PhysRevA.44.5913}.

It is clear from \cref{eq:cat} that at multiples of $e \simeq 2\pi |z|$ the cat becomes approximately coherent at $\ket{\pm z}$ again, see \cref{fig:wigner}, depending on whether the multiple is odd or even.
At these points the classical Rabi oscillations reappear: these are the revivals.
For the even-numbered revivals, the state approximately returns to the original coherent state $\ket{z}$, which is why (an approximation to) these revivals can be seen by including only loop effects.
For the odd-numbered revivals, however, the state is approximately coherent, but at $\ket{-z}$, meaning it has undergone a phase shift of $\pi$ relative to the initial state $\ket{z}$: this is back-reaction on the field, and indeed is as severe a back-reaction as is possible with a single undamped mode.
Since $\ket{z}$ has significant Fock space components up to $n \approx \mathrm{e} |z|^2$, capturing this back-reaction requires the emission of a large large number, $\mathcal{O}(\mathrm{e} |z|^2)$, of photons, as we saw above in \cref{fig:second-revival}.

These results have direct implication for models of back-reaction in QED.
It has been suggested~\cite{Ilderton:2017xbj} to consider transitions between \emph{different} initial and final coherent states $\ket{z_i}$ and $\ket{z_f}$ to account for back-reaction in the form of depletion.
This essentially amounts to displacing $a \mapsto a + z_i, a^\dagger \mapsto a^\dagger + \zc_f$, so that the gauge field $A_\mu$ is shifted by a complex value.
Applying this with $z_f = -z_i$, one can capture the odd-numbered revivals in JCM, without including emissions, but the even-numbered revivals would be missed.
Further, between revivals, we have seen that back-reaction puts the field in a non-classical cat state, and no single $\ket{z_f}$ offers a good approximation without emissions.
Instead, the state \cref{eq:cat} and \cref{fig:wigner} suggest that to go beyond~\cite{Ilderton:2017xbj} one should consider a superposition of coherent states, or other non-classical states, to better capture quantum aspects of back-reaction.

What we have discussed here is an example of back-reaction at strong coupling, since $e$ is required to be large for collapse/revivals to occur at all.
In the next section we consider a situation more analogous to QED in which the coupling is kept fixed and small, but the strength of the external field $a_0$ is allowed to vary.

\section{Back-reaction at high intensity and the Furry expansion}
\label{sec:furry}
We turn now to the Furry picture proper, where interactions with the background field are taken into account to all orders in $a_0$, but emissions and loops are still treated perturbatively as a series in the coupling $e$, which is therefore now assumed to be \emph{small}.

It has been conjectured in QED that, in a high-intensity background, the effective coupling is not the fine-structure constant $\alpha$, but rather $\alpha \chi^{2/3}$, where the `quantum nonlinearity parameter' $\chi$ is
\begin{equation}
    \chi = \frac{e}{m^3} \sqrt{(F^{\mu\nu} k_\nu)^2}
    \;,
    \label{eq:chi-def}
\end{equation}
in which $e$ is the elementary charge, $m$ the electron mass, $F^{\mu\nu}$ the background field strength, and $k_\mu$ a probe particle momentum.
$\chi$ is essentially the product of particle energy and field intensity $a_0$.
Such a dependence would imply a breakdown of Furry picture perturbation theory at $a_0 \gg 1$, in the sense that all loop corrections would have to be resummed.
This conjecture follows from the scaling, with intensity or $\chi$, of some higher loop diagrams calculated in constant plane wave fields.
Such calculations are extremely challenging, even more so in more general fields.
While the locally constant field approximation would suggest that the same scaling applies to all fields (even beyond plane wave~\cite{Nikishov:1964zza}), there are exactable solvable examples which show that the scaling does not hold for all fields~\cite{Ilderton:2019vot}, and it does not hold if the composite parameter $\chi$ is made large by going to high energy~\cite{PhysRevD.99.085002,PhysRevD.99.076004}.

In the much simpler JCM, we will  be able determine the asymptotic scalings of all loop diagrams, with any number of emissions.
While we have no energy parameter in JCM, and thus no $\chi$, we do have a coupling/field strength $a_0$ analogous to that in QED; our interest here is therefore in the asymptotic scaling of diagrams at high $a_0$.

Some low-loop-order amplitudes can be explicitly calculated by `brute force' expansion of the $S$-matrix, keeping track of powers of $e$ and $z$.
For example, writing $\xi = a_0 \mathrm{e}^{i\phi}$, we can find the tree-level amplitude for spin flip without emission, from above,
\begin{equation}
    \begin{split}
        \downarrow
        \raisebox{-1pt}{\includegraphics{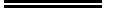}}
        \uparrow \;
        & =
        \bra{0, \uparrow} \dispSmatrix \ket{0, \downarrow}^{(0)}
        \\
        & =
        -i \mathrm{e}^{i\phi} \sin a_0
        \label{eq:forward}
        ,
    \end{split}
\end{equation}
its order $e^2$, or $1$-loop, correction,
\begin{equation}
    \begin{split}
       \downarrow
       \includegraphics{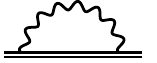}
       \uparrow
       \;
       & =
       \bra{0, \uparrow} \dispSmatrix \ket{0, \downarrow}^{(1)}
       \\
       & =
       e^2
       \frac{i \mathrm{e}^{i\phi}}{8}
       \left(\sin a_0
           -\frac{\cos a_0}{a_0} + \frac{\sin a_0}{a_0^2}
       \right)
       ;
    \end{split}
\end{equation}
and the tree-level amplitude for flip with one emission, which is the analogue of non-linear Compton scattering~\cite{Nikishov:1964zza} (for a review see~\cite{Seipt:2017ckc}),
\begin{equation}
    \begin{split}
        \downarrow
        \raisebox{0.5pt}{\includegraphics{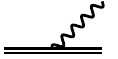}}
        \uparrow
        \;
        & =
        \bra{1, \uparrow} \dispSmatrix \ket{0, \downarrow}^{(0)}
        \\
        & =
        - e \frac{i \mathrm{e}^{2i\phi}}{2}
        \left( \cos a_0 - \frac{\sin a_0}{a_0}\right)
        \label{eq:nlc}
        .
    \end{split}
\end{equation}
(\cref{fig:flip-e-2} shows the flip probability based on the above results; this will be discussed in \cref{sect:inc}.) In contrast to QED, these amplitudes are \emph{bounded} for large $a_0$; we will show below that this holds to all loop orders.

\subsection{Furry expansion to all loop orders}
\label{sec:all-orders}
For concreteness, we consider the $S$-matrix element for spin flip with the emission of $n$ photons (to all loops), $\bra{n,\uparrow} \dispSmatrix \ket{0,\downarrow}$, but the arguments in this section go through regardless of incoming/outgoing spin state.
As in field theory, we write the $S$-matrix in normal-ordered form, using the formula~\cite{Blasiak1,Blasiak2}
\begin{equation}
    (a^\dagger a)^r = \sum_{k = 0}^r \stirling{r}{k} (a^\dagger)^k a^k
    \label{eq:normal-ordering}
\end{equation}
where $\textstyle\stirling{r}{k}$ is a Stirling number of the second kind.
Applying this to the $S$-matrix and using the recurrence relation defining the Stirling numbers, we find
\begin{widetext}
    \begin{equation}
        \bra{\uparrow} \dispSmatrix \ket{\downarrow}
        =
        -i \sum_{r = 0}^\infty \sum_{k = 0}^\infty
        \frac{(-1)^r e^{2r+1-2k} }{(2r+1)!}
        \stirling{r + 1}{k + 1}(ea^\dagger + \bar\xi)^k (ea + \xi)^{k+1}
        .
        \label{eq:amp-furry-start}
    \end{equation}
\end{widetext}
This is the analogue of a sum over Feynman diagrams with $2r + 1$ vertices and $k + 1$ incoming ($k$ outgoing) photon legs, each of which can be coupled either to the background or to an absorbed (emitted) photon.
In the former case, the diagram picks up a factor $a_0$ (up to a phase), and in the latter a factor $e$.
A spin-flip diagram with $2r + 1$ vertices and $2k + 1$ external lines has $\ell := r - k$
contractions, or loops, and with $n$ emitted photons will be of order $e^{2\ell + n}$.
Performing the sum at fixed $\ell$ then gives us the $\ell$-loop contribution,  with all orders in $a_0$ included; this is the Furry loop expansion.
The Stirling number $\stirling{r + 1}{r + 1- \ell}$ is then the analogue of a loop integral, as it arises from contractions in normal-ordering the $S$-matrix.

We therefore go over to variables $r$ and $\ell$, and perform the sum in \cref{eq:amp-furry-start} at fixed $\ell$.
To do so we require the photon contribution to the $S$-matrix element, viz.
\begin{multline}
    \bra{n} (ea^\dagger + \bar\xi)^{r-\ell} (ea + \xi)^{r - \ell +1} \ket{0}
    \\
    =
    \sqrt{n!} \binom{r-\ell}{n} e^n {\bar\xi}^{r-\ell-n}\xi^{r-\ell+1} \;.
    \label{eq:photonic}
\end{multline}
Thus, expressing the $\ell$-loop amplitude in terms of $a_0$ and $e$, we have
\begin{widetext}
    \begin{equation}
        \bra{n, \uparrow} \dispSmatrix \ket{0, \downarrow}^{(\ell)}
        =
        \;
        \downarrow
        \raisebox{-34pt}{\includegraphics{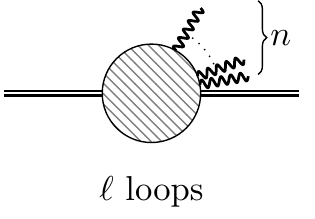}}
        \uparrow
        \;
        =
        -i \sqrt{n!} \, \mathrm{e}^{i(n+1)\phi}
        \bigg(\frac{e}{a_0}\bigg)^{2\ell + n}
        a_0
        \sum_{r = 0}^\infty
        P_{\ell, n}(r) \frac{(-1)^r a_0^{2r}}{(2r + 1)!} \;,
        \label{eq:amp-furry}
    \end{equation}
\end{widetext}
in which
\begin{equation}
    P_{\ell, n}(r) := \stirling{r + 1}{r + 1 - \ell} \binom{r - \ell}{n} \;.
    \label{eq:PLN-DEF}
\end{equation}
The properties of the binomial coefficient and the Stirling numbers mean that $P_{\ell,n}(r)$ is a polynomial in $r$; we will demonstrate this below, here we just observe that we may shift $P_{\ell,n}$ outside the sum by trading each power of $r$ in the polynomial for the same power of
\begin{equation}
	D := \frac{a_0}{2}\frac{\partial}{\partial a_0} \;,
\end{equation}
acting on $a_0^r$ in (\ref{eq:amp-furry}).
Doing so allows us to perform the sum over $r$, yielding
\begin{multline}
    \bra{n, \uparrow} \dispSmatrix \ket{0, \downarrow}^{(\ell)}
    =
    \\
    -i \sqrt{n!} \mathrm{e}^{i(n+1)\phi} \left(\frac{e}{a_0}\right)^{2\ell + n} a_0
    P_{\ell, n}(D) \frac{\sin a_0}{a_0}.
    \label{eq:spin-flip-furry-final}
\end{multline}
This is the $\ell$-loop Furry picture spin flip amplitude with the emission of $n$ photons.
We do not believe this expression has previously appeared in the JCM literature.
It remains to ascertain the leading-order behaviour of the amplitudes as a function of $a_0$.
To do so we need some properties of $P_{\ell, n}$.

In the definition (\ref{eq:PLN-DEF}) the binomial coefficient is a polynomial in $r$ of degree $n$.
That the Stirling number is also a polynomial in $r$ follows from its  defining recurrence relation, $\stirling{r}{r} = 1$ and
\begin{equation}
    \stirling{r + 1}{r + 1 - \ell} = \sum_{k = 0}^r k \stirling{k}{k - \ell - 1} \;, \quad \ell \ge 1
    \;.
    \label{eq:stirling-rec}
\end{equation}
For $\ell = 1$, \cref{eq:stirling-rec} is a sum of linear terms, so $\stirling{r+1}{r+\ell-1}$ is a quadratic in $r$.
For $\ell = 2$,  \cref{eq:stirling-rec} is a sum of cubics (and lower order terms), so $\stirling{r+1}{r+\ell-1}$ is a quartic in $r$ and so on.
Thus, $\stirling{r + 1}{r + 1 - \ell}$ is in general a polynomial of degree $2\ell$ in $r$ and $P_{\ell,n}(r)$ is a polynomial of degree $2\ell + n$ in $r$.

Now, the binomial part of $P_{\ell,n}$ has highest-power coefficient $1/n!$.
Because $\sum_{k = 0}^r k^p = r^{p+1}/(p+1) + \mathcal{O}(r^p)$, the highest-power coefficient in $\stirling{r+1}{r+1-\ell}$ is $1/(2\ell)!! = 1/(2^\ell \ell!)$.
Thus we have the highest-power behaviour\footnote{%
    The full polynomial may be determined by fitting to $\stirling{r + 1}{r + 1 - \ell}$ for $2\ell + 1$ values of $r$, or recursively using \cref{eq:stirling-rec} and Faulhaber's formula.
    One may check that $P_{0,0} = 1$, $P_{1,0}(r) = r(r+1)/2$, and $P_{0,1}(r) = r$ which recovers the one-loop amplitudes above.
    }
\begin{equation}
    P_{\ell,n}(D) = \frac{D^{2\ell + n}}{2^\ell n! \ell!}  + \ldots
    \label{eq:leading-coeff}
\end{equation}
We can finally give the leading-order behaviour in $a_0$.
The highest power of $a_0$ comes from the highest-power term in $P_{\ell,n}(D)$, which is of order $a_0^{2\ell + n}$, precisely the inverse of the prefector.
The derivatives in $D^{2\ell+n}$ act on $\sin(a_0)/a_0$; the leading-order behaviour is given by that term in which all derivatives act on $\sin(a_0)$.
We conclude that, in JCM, $\ell$-loop Furry picture diagrams with $n$ emitted photons have the leading-order large $a_0$ behaviour
\begin{equation}
    \sim
    \frac{e^n}{2^n n!}
    \frac{e^{2\ell}}
    {8^\ell \ell!}
        \begin{Bmatrix}
            \sin a_0 \\
            \cos a_0
        \end{Bmatrix}
+ \mathcal{O}(a^{-1}_0)
\end{equation}
with either $\sin$ or $\cos$ depending on the diagram in question.
There is no power-law scaling with $a_0$.
This can be contrasted with the conjectured behaviour of QED amplitudes at high intensity~\cite{PhysRevD.21.1176,Fedotov_2017,PhysRevD.99.085002,PhysRevD.99.076004}, where successive loop orders come with higher powers of $a_0$.
In JCM however, the effective expansion parameter does not acquire an intensity dependence; it clearly remains $e$ in the Furry picture.
Our arguments extend to amplitudes for no spin flip, for other initial spin states, and (see the appendix) to arbitrary detuning.

\subsection{Inclusive vs. exclusive observables at weak coupling}\label{sect:inc}
Consider the regime of weak coupling, $e<1$, and strong field $a_0>1$, which mirrors the typical situation in QED.
    For $e<1$ there can be no collapse or revival, and so the tree-level background field approximation (Rabi oscillations) is a good approximation to the full physics of JCM.
    Despite this, it is interesting to note the relative importance of loops and emissions -- in the context of $\alpha\chi^{2/3}$, only loop corrections are studied, whereas physical observables are at least partially inclusive~\cite{Yennie:1961ad,Lavelle:2005bt}.

\begin{figure}[t!]
    \centering
    \includegraphics[width=\linewidth]{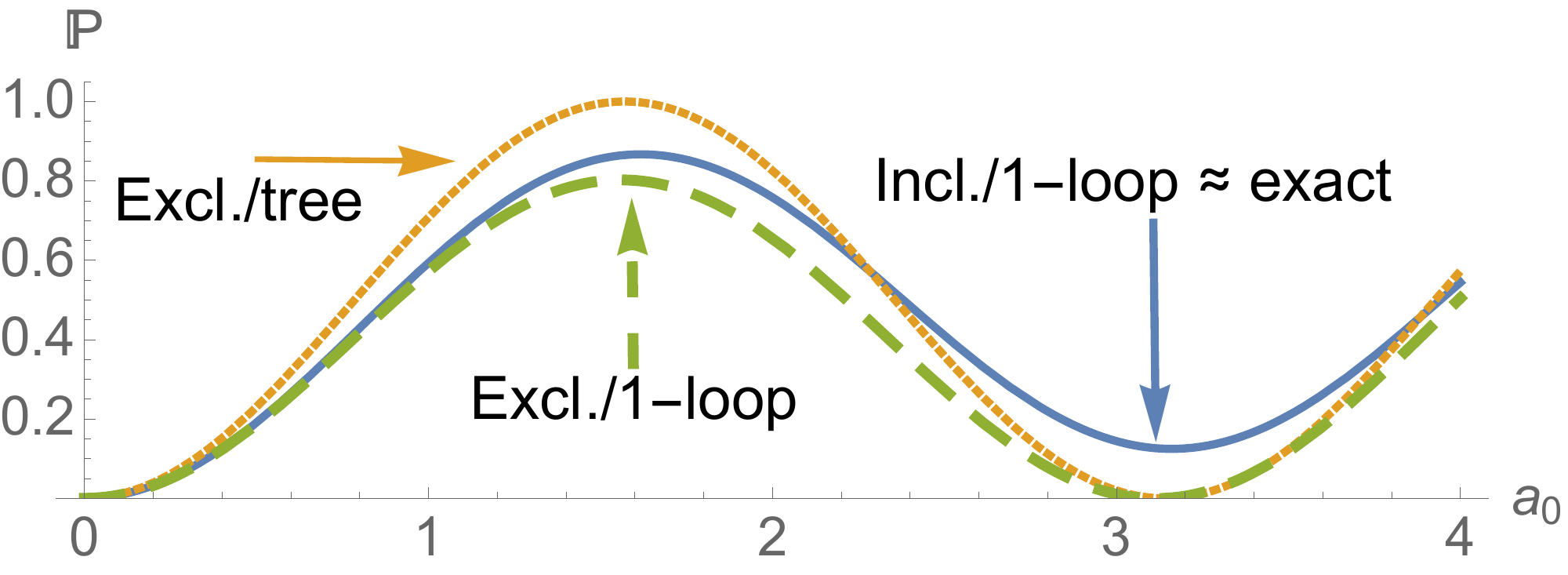}
    \caption{%
        Spin flip probability at fixed coupling $e=0.75$.
        As intensity increases, 1-loop corrections are needed to capture the exact result.
        For higher intensities, though, emission (non-linear Compton scattering) is also required, i.e.~the exclusive and inclusive probabilities differ.
        At this coupling the order $e^2$ inclusive probability is almost equal to the exact result.
        \label{fig:flip-e-2}
    }
\end{figure}

In \cref{fig:flip-e-2} we plot the spin flip probability in several approximations at $e=0.75$, using the amplitudes~\crefrange{eq:forward}{eq:nlc}.
For low field strength $a_0$, the tree level exclusive probability (Rabi oscillations) is a good approximation.
For $a_0\simeq 1$, though, its 1-loop correction is needed to track the exact result.
As $a_0$ increases, we see that the exclusive probability differs from the inclusive, which includes 1-photon emission.
Hence for some $a_0$ spin flip occurs predominantly through non-linear Compton scattering.
This suggests that, even for weak coupling, one still has to account for emission as well as loops.

\section{Conclusions}
\label{sec:conclusion}

We have studied the background field approximation, and corrections to it, in the Jaynes-Cummings model (JCM), as a toy model of QED.
The simplicity of the model allows us to calculate exactly in all parameters, and this has yielded new insights of relevance to both JCM and QED.

We have seen that even (the analogue of) all-orders loop corrections in JCM are insufficient to capture collapse/revival physics.
In particular, odd-numbered revivals occur when the field has experienced the maximal back-reaction available in the model (a $\pi$ phase shift).
Without back-reaction in the form of significant photon emission, the revivals are not seen.

It is not strictly possible in JCM to consider back-reaction in the form of \emph{depletion}, since the Hamiltonian can change photon number by at most $\pm 1$.
As an extension which would allow for depletion, we could add a second photon mode with the same frequency, taking the interaction to be $H_I = (g_1 a_1 + g_2 a_2) \tau_+ + \operatorname{h.c.}$ with couplings $g_1$ and $g_2$.
An  $SU(2)$ rotation decouples one of the modes, and we could proceed as above~\cite{PhysRevA.67.053801}.
However, rotating back to the original representation, the analytic $S$-matrix becomes unmanagable for large inital photon numbers~\cite{PhysRevA.67.053801},
and numerics become expensive because of the greatly enlarged state space.
In principle, though, our methods could be applied to look at, e.g., how an initial mode-$1$ coherent state depletes into mode-$2$ photons.

We have learnt several lessons for QED.
For example~\cite{Ilderton:2017xbj} attempts to model back-reaction by allowing an initially coherent state to evolve into another; we have seen that to improve upon this one could instead use non-classical superpositions of coherent states, because in JCM the system is, between revivals, in a cat state, exhibiting the quantum nature of back-reaction.

We have also considered the Furry expansion of amplitudes in JCM, prompted by the conjectured QED behaviour that the effective coupling becomes intensity-dependent~\cite{PhysRevD.21.1176}.
We have seen that there is no such behaviour in JCM.
The implications for QED are not conclusive.
In particular, JCM has only a finite number of degrees of freedom, and `loop' contributions do not involve momentum integrals.
There is also no energy variable, so that while we can identify an intensity parameter corresponding to $a_0$ in QED, there is no composite parameter like $\chi$; it may be, due to the lack of universal $\chi$-dependence in QED~\cite{Dinu:2013gaa,Gies:2014jia,Dinu:2015aci,PhysRevD.99.076004,PhysRevD.99.085002}, that the precise way in which $\chi$ enters loop integrals is important.
It would in future work be interesting to find a more complex, but still solvable, model, which brings the calculations here closer to QED proper.

Encouragingly, though, our results mean that the conjectured breakdown of background field perturbation theory at high intensity in QED is \emph{not} a necessary feature of general quantum theories in background fields.

\acknowledgements

\emph{We thank Tom Heinzl and Daniel Seipt for useful discussions.
The authors are supported by the Leverhulme Trust, grant RPG-2019-148.}

\appendix

\section{Loop expansion with non-zero detuning}

\newcommand{\Deltilde}{\ensuremath{\Delta}}

A non-zero detuning does not change the conclusion of \cref{sec:furry}; the same argument goes through, with a few extra steps that obfuscate the presentation.
We will illustrate using, as before, the $S$-matrix element for spin flip, which is~\cite{GerryKnight}
\begin{equation}
    \bra{\uparrow} \mathcal{S} \ket{\downarrow}
    =
    -\frac{ia}{\sqrt{\Deltilde^2 + \hat{n}}} \sin e \sqrt{\Deltilde^2 + \hat{n}}
\end{equation}
Here we have defined the \emph{dimensionless} detuning $\Deltilde = (\omega - \omega_a)/2|g|$, in the notation used in \cref{eq:JC-hamiltonian}.
As for zero detuning, we expand the sine in its power series, displace $ea \mapsto ea + \xi$, normal order using \cref{eq:normal-ordering}, identify a variable $\ell$ corresponding to the number of loops, and take the initial and final photonic states to be $\ket{0}$ and $\ket{n}$, respectively, using \eqref{eq:photonic}.
After using the binomial theorem and a change of summation variables, we arrive at
\begin{widetext}
    \begin{equation}
        \mathcal{M}_\ell
        :=
        \frac{\mathrm{e}^{i\theta}}{\sqrt{n!}} \bra{n, \uparrow} \dispSmatrix \ket{0, \downarrow}^{(\ell)}
        =
        \left(\frac{e}{a_0}\right)^{2\ell + n} a_0
        \sum_{s = 0}^\infty \sum_{r = 0}^\infty
        \frac{(-1)^{s+r}}{(2s + 2r + 1)!} \binom{s+r}{s} (e\Deltilde)^{2s} e^{2\ell + n} \stirling{r + 1}{r - \ell + 1} \binom{r-\ell}{n} a_0^{2r}
    \end{equation}
    for a phase $\theta$ which is unimportant for our purposes.
    The effect of non-zero detuning is that in any loop diagram, any number of ``two-point vertices'' $\propto (e\Deltilde)^2$ can be inserted.
    Intuitively, this cannot make a diagram grow faster with $a_0$.

    As before, we identify
    $
    P_{\ell,n,s} = \binom{s+r}{s} \binom{r-\ell}{n} \stirling{r-\ell+1}{r+1}
    $
    as a polynomial in $r$, of degree $2\ell + n + s$.
    Therefore,
    \begin{align}
        \mathcal{M}_\ell
        =
        \left(\frac{e}{a_0}\right)^{2\ell + n} a_0
        \sum_{s = 0}^\infty
        P_{\ell, n, s}(D)
        (e\Deltilde)^{2s}
        \sum_{r=0}^\infty
        \frac{(-1)^{s+r}}{(2s+2r+1)!} a_0^{2r}
        =
        \left(\frac{e}{a_0}\right)^{2\ell + n}
        a_0 \sum_{s = 0}^\infty
        P_{\ell, n, s}(D)
        \left(
            \frac{e\Deltilde}{a_0}
        \right)^{2s}
        \sinc_{\ge s}(a_0)
    \end{align}
\end{widetext}
where by $\sinc_{\ge s}$ we mean $\sinc$ but with the $s$ first terms of its power series omitted.
The $s = 0$ term recovers the zero-detuning result, i.e., it goes like $\sin(a_0)$ or $\cos(a_0)$ plus terms that are $\mathcal{O}(a_0^{-1})$.
By power-counting, the terms with $s \ge 1$ are $\mathcal{O}(a_0^{-2})$.
Again, other initial and final spin states can be handled in the same way.

\bibliography{JCF-bib}

\end{document}